\documentclass[aps,prb,preprint,groupedaddress]{revtex4}

\usepackage{graphicx}
\usepackage{marvosym}

\begin{document}

\title{Theoretical characterization of the electronic properties of
extended thienylenevinylene oligomers}

\author{C. Krzeminski, C. Delerue$^{*}$, G. Allan, V. Haguet and D. Sti\'evenard
}

\affiliation{Institut d'Electronique et de Micro\'electronique du Nord, D\'epartement Institut Sup\'erieur d'Electronique du Nord, 41 Boulevard Vauban, 59046 Lille C\'edex, France}

\author{P. Fr\`ere, E. Levillain and J. Roncali}

\affiliation{Ing\'enierie Mol\'eculaire et Mat\'eriaux Organiques, Universit\'e d'Angers, 2 Bd Lavoisier, 49045 Angers, France}

\email[*]{delerue@isen.fr}

\vspace{10cm}

\begin{abstract}
We present semi-empirical tight binding calculations on thienylenevinylene oligomers up to the hexadecamer stage ($n=16$) and {\it ab initio} calculations based on the local density approximation up to $n=8$. The results correctly describe the experimental variations of the gap versus size, the optical spectra and the electrochemical redox potentials. We propose a simple model to deduce from the band structure of the polymer chain the electronic states of the oligomers close to the gap. We analyze the evolution of the gap as a function of the torsion angle between consecutive cells: the modifications are found to be small up to a $\sim$30$^{\circ}$ angle. We show that these oligomers possess extensive $\pi$-electron delocalization along the molecular backbone which makes them interesting for future electronic applications such as molecular wires.

\end{abstract}


\maketitle

\section{Introduction}

In the microelectronic road map, one approach for the conception of nanodevices is the "bottom-up" one which consists of arranging individual atoms or molecules to get a given function. This kind of approach has been strongly developed thanks to the scanning tunneling microscope which offers the possibility of both observation and manipulation of single atoms or molecules (for a recent review, see \cite{ref1}). Among the infinite possibilities of synthetic organic chemistry, linearly $\pi$-conjugated systems with multinanometer dimensions offer very exciting opportunities due to their potential use as molecular wires in nanoelectronic devices. 

Such a purpose requires stable linear $\pi$-conjugated systems combining optimal $\pi$-electron delocalization with approaching the present limit of nanolithographic techniques, i.e. the 100 \AA \  regime. In this context extended $\pi$-conjugated oligomers based on phenyleneethynylene and thiopheneethynylene have been proposed as potential molecular wires \cite{ref2,ref3}. However, all these structures exhibit a more or less rapid saturation of the effective conjugation length and hence of the HOMO-LUMO gap due to excessive $\pi$-electron confinement related to structural factors such as aromatic resonance energy or as rotational disorder \cite{ref4}. Consequently, the width of the gap converge towards a limiting value for chain lengths significantly shorter than a maximum chain dimension.

 Although the first members of oligomers of the thienylenevinylene oligomers ({\it n}TVs) have been known for several years \cite{ref5,ref6}, their lack of solubility did not allow neither the synthesis of longer oligomers nor the detailed analysis of their electronic and electrochemical properties. As shown recently, the introduction of solubilizing alkyl chains at the 3 or 3 and 4 positions of the thiophene allows the synthesis of highly extended {\it n}TVs (up to the hexadecamer stage, $n=16$) with chain dimensions close to the present limits of the nanolithography ($\sim$ 10 nm) \cite{ref7}. Owing to their recent emergence as a new class of extensively $\pi$-conjugated oligomers, {\it n}TVs have not been subject to theoretical works, in sharp contrast with parent series of conjugated oligomers such as oligothiophenes, oligophenylenevinylenes or oligoanilinies \cite{ref8,ref9,ref10}. 

This work presents a theoretical analysis of the chain length dependence of the electronic properties of nTVs involving one to 16 thiophene rings. We discuss the nature of the states, the optical properties and the ionization energies of the oligomers and the theoretical results are then compared to the available experimental data.

\section{CALCULATIONS}

The atomic and electronic structure of molecules or solids can be calculated by various techniques which rely on different degrees of approximation. {\it Ab initio} configuration interaction techniques are extremely powerful to predict spectroscopic properties but cannot be applied to molecules containing a large number of atoms like nTVs. Thus we have performed {\it ab initio} calculations based on the local density approximation (LDA) \cite{ref11} using the DMOL code \cite{ref12}. For the computation, we use a double numerical basis set (two atomic orbitals for each occupied orbital in the free atom) together with polarization functions ({\it 2p} for H and {\it 3d} for S and C). All the results presented in this paper are based on the spin-density functional of Vosko {\it et al} \cite{ref13}. Results obtained with another functional \cite{ref14} show only minor differences ($\sim$ 2 meV on the gap of thiophene). To save computational workspace, the hexyl groups are replaced by methyl groups (\ref{fig:fig1}). We check that this does not change the nature and the energy of the electronic states of interest, i.e. close to the gap. For thiophene, 2TV and 3TV, an optimization of the geometry is performed by minimization of the total energy with respect to the atomic coordinates \cite{ref12}. We obtain that the molecules are almost planar (this point is discussed in the next section) and that the geometry of the different groups are almost identical. Thus, for longer {\it n}TVs $(n>3)$, we have just replicated identical cells to build planar molecules without optimization of the geometry. Because of computational limits, we are able to apply LDA calculations only up to $n=8$. For longer chains, a simpler approach is required. This is achieved by using a self-consistent tight binding calculation which is a semi-empirical technique close to the simple H\"{u}ckel theory. C and S atoms are represented by one $s$ and three $p$ atomic orbitals, and H atoms by one $s$ orbital. The elements of the Hamiltonian matrix, restricted to first nearest neighbor interactions and to two center integrals, are written in terms of a small number of parameters (15) following the rules of  ref. \cite{ref15}. First nearest neighbor interactions depend on the interatomic distance following Harrison\'s rules \cite{ref16}. For the sake of consistency with LDA calculations, we apply the following procedure to determine the parameters. First, we calculate in LDA the electronic structure of simple molecules: methane, ethane, ethylene, acethylene and CS$_{2}$. Then the parameters are determined by a fit of the LDA electronic structure using a least square minimization technique. The charge transfers between the atoms are also adjusted on those obtained in LDA using a Mulliken analysis \cite{ref12,ref17}. The calculation is made self-consistent by adding to the diagonal matrix elements charge dependent Coulomb terms \cite{ref18}. Details on the calculations and on the tight binding parameters are given in Appendix. 

Let us discuss the respective limits of LDA and tight binding calculations. LDA techniques are very powerful to predict ground state properties of molecules or solids \cite{ref19}  but, strictly speaking, are not applicable to excited states. Actually, one-electron gaps calculated in LDA are usually too small compared to experimental optical gaps \cite{ref19,ref20}. Thus a correction is usually applied to LDA gaps corresponding to a rigid shift of the unoccupied states with respect to occupied ones (the so-called scissor operator). The correction results from a change of the exchange-correlation potential through the gap (correlation effects) and also, for finite systems like molecules, from electrostatic effects and Coulomb interaction between the electron and the hole in the excited state \cite{ref20,ref21}. It is often assumed that this correction does not depend too much on the size of the chain, i.e. that the variations of the gap are well predicted in LDA \cite{ref21}. Recent calculations on the gap of silicon nanocrystallites tend to confirm this statement \cite{ref21}. Finally, the same underestimation of the gap is expected here in tight binding because the parameters are adjusted on the electronic structure of small molecules calculated in LDA.

\section{RESULTS AND DISCUSSION}

The synthesis and characterization of the optical and electrochemical properties of the oligomers have been described elsewhere \cite{ref7}. Fig. \ref{fig:fig2} reproduces typical UV-visible absorption spectra recorded in CH$_{2}$Cl$_{2}$ for 4TV, 6TV, 8TV, 12TV and 16TV. In each case, the spectrum exhibits a well-resolved vibronic fine structure indicative of the persistence of a relatively planar and rigid geometry even for longest chains. From the spectra we define the optical gap by the maximum of the lowest transition which we report on Fig. \ref{fig:fig3} versus $N_{c}$ the number of sp$^{2}$ carbons in the molecule ($N_{c}=6n-2$). Chain extension leads to a shift of the maximum absorption corresponding to a narrowing of the gap. It reaches 1.95 eV for $N_{c}=94 \ (n=16)$, the smallest ever reported value for a $\pi$-conjugated oligomer of homogeneous chemical structure. A representation of the gap with respect to $1/N_{c}$ (or 1/n) is common in the literature \cite{ref23}  but in our case it is a good approximation only for small chains because the gaps saturate for large N$_{c}$ \cite{ref7}. A saturation, corresponding to a limit of the effective conjugation length, is reported for all the series of oligomers studied up to now \cite{ref24}. Meier {\it et al.} \cite{ref24} have suggested that this limit is reached when the bathochrom shift (or red shift) of $\lambda_{max}$ resulting from the lengthening of the chain is smaller than 1 nm. The shift of 6 nm that we measure between 12TV and 16TV shows that the limit is close to $n=16 $but it is not yet reached. The comparison with other oligomer series shows that {\it n}TVs have the longest limit of effective conjugation as well as the smallest gap \cite{ref7}. It remains to compare the variations of the gap with those predicted by the theory. 

Gaps calculated in LDA and in tight binding are plotted in Fig. \ref{fig:fig3}. The results obtained by the two techniques are reasonably close which is remarkable considering the simplicity of the tight binding calculation. As discussed above, we need to take into account the underestimation of the gap by the LDA. Applying to the theoretical gaps an upward shift of 1.2 eV (Fig. \ref{fig:fig3}), we obtain a correct agreement between theory and experiments (within $\sim$0.2 eV): thus the experimental variations of the gap are well described by the theory. 1.2 eV is a reasonable and common value for the gap correction \cite{ref20}. The tight binding gap is well approximated in eV by $1.85 + \frac{14.96} {N_{c}^{1.12}}$ including the shift of 1.2 eV. The limit (1.85 eV) is the band gap energy of the polymer (polythienylenevinylene) calculated in tight binding (+ 1.2 eV), in good agreement with the experimental values of the order of 1.80 eV \cite{ref25}. An exponent (1.12) between 1 and 2 is commonly obtained in semiconductor quantum dots \cite{ref26}. 

Fig. \ref{fig:fig4} shows the densities of LDA eigenstates from thiophene to 6TV. Similar results are obtained in tight binding. We see that the evolution of the HOMO and LUMO states is almost symmetrical which implies an efficient delocalization of both states. This is confirmed by the analysis of the molecular orbitals showing that these electronic states mainly derive from the interactions between $\pi$ orbitals along the whole molecule. These results show that in principle efficient electron or hole conductivity could be achieved in {\it n}TVs providing that the {\it n} or {\it p}-type doping is realized (but to conclude the effect of the charge carriers on the conformation of the molecule should be analyzed). 

Fig. \ref{fig:fig5} shows the electronic band structure of the polymer chain calculated in tight binding. The highest occupied band and the lowest unoccupied one  have a large dispersion in energy which is consistent with an efficient delocalization of the states. Interestingly, the energies of these conduction $\epsilon_{e}$ and valence $\epsilon_{h}$  bands are reasonably well represented by simple cosine laws:

\begin{equation}
\epsilon_{e} = 3.65-0.89 \cos (ka/ 2 ) 
\label{eq:eq_one}
\end{equation}

\begin{equation}
\epsilon_{h} = 1.13 + 0.98 \cos (ka/ 2 )
\label{eq:eq_two}
\end{equation}

where $k$ is the wave vector and $a$ is the length of the translational unit cell which contains two thienylenevinylene groups (the energies are given in eV and they do not include the shift of 1.2 eV). The formula (1) and (2) are equivalent to the tight binding dispersion laws of a linear chain with one orbital in a unit cell of length $a/2$ (corresponding here to one orbital in each thienylenevinylene group) and with a constant coupling only between nearest neighbor orbitals. From this we can deduce a simple way to calculate the electronic states of the oligomers close to the gap. The main assumption is that the two bands are independent. Then the electronic states of {\it n}TV are modeled by the eigenstates of the Hamiltonian of the finite linear chain containing $n$ orbitals. This is equivalent to the well-known analytical H\"uckel theory for polyenes \cite{ref27}. The solutions of such problem being analytic, we obtain immediately:

\begin{equation}
\epsilon_{(e_{p})} = 3.65-0.89 \cos (\frac{p \pi}{n+1})
\label{eq:eq_three}
\end{equation}

\begin{equation}
\epsilon_{(h_{p})} = 1.13+0.98 \cos (\frac{p \pi}{n+1})
\label{eq:eq_four}
\end{equation}

where $e_{p}$ ($h_{p}$) represents the electron (hole) states and $p$ is an integer index such that $ 1 \leq p < n$ . Some electron and hole states ($p = 1, 2, 3$) of 6TV calculated with eqn. (\ref{eq:eq_three}) and (\ref{eq:eq_four}) are represented in Fig. \ref{fig:fig5} (horizontal dashed lines): they agree relatively well with those obtained with the full tight binding calculation (box on the right side of Fig. \ref{fig:fig5}). The agreement is mainly qualitative because many effects are left out in the model such as the coupling between the two bands. But this kind of simple model can be particularly helpful to investigate different problems related to (like optical properties discussed below). 

We have also calculated the optical absorption of {\it n}TVs in tight binding. Working within the so-called dipole approximation, the intensity of the absorption is proportional to the momentum matrix elements between initial and final electronic states \cite{ref28}. The optical matrix elements are computed like in ref. \cite{ref29}. Phonon-assisted transitions are not considered in our calculations. Fig. \ref{fig:fig2} shows that the calculated transitions agree relatively well with the experiments: in particular we can explain the origin of the two broad bands in the absorption spectra (however the energy of the upper band seems to be slightly underestimated by the calculations for large {\it n}. Note that this agreement is obtained when a correction of 1.2 eV is applied to the gap. Each absorption band is indeed composed of several transitions whose number increases with {\it n}. The most intense transitions, identified by labels in \ref{fig:fig2} and by arrows in \ref{fig:fig5} in the particular case of 6TV, can be easily interpreted using the results of the analytical H\"{u}ckel theory for polyenes \cite{ref27}. The lowest absorption band is composed of transitions $h_{p} \rightarrow  e_{p}$. The transitions $h_{p} \rightarrow e_{m}$ with $p \ne m$ are not visible in Fig. \ref{fig:fig2}: they are almost dipole forbidden for the following reasons. The eigenfunctions $|\Psi_{p}^{h}>$ and $|\psi_{m}^{e}>$ corresponding respectively to the states h$_{p}$ and e$_{m}$ can be written:

\begin{equation}
|\psi_{p}^{h}=\sum_{i=1,n}a_{i}^{p}|\phi_{i}^{h}>
\label{eq:eq_five}
\end{equation}

\begin{equation}
|\psi_{m}^{e}=\sum_{i=1,n}a_{i}^{m}|\phi_{i}^{e}>
\label{eq:eq_six}
\end{equation}

where {\it i} is the index of the thienylenevinylene group. $|\phi_{i}^{h}>$ and $|\phi_{i}^{e}>$  are the effective orbitals on the site {\it i} corresponding respectively to the valence {\it (h)} and conduction {\it (e)} bands. It is important to point out that the coefficients ($a_{i}^{p}$ or $a_{i}^{m}$) in eqn. (5) and (6) do not depend on the band ({\it e} or {\it h}) because, in the case of a linear chain of identical orbitals with a constant coupling between the nearest neighbors, the eigenstates do not depend on the strength of the coupling. To characterize the transition $h_{p} \rightarrow  e_{m}$, we must write the matrix element of the momentum:

\begin{equation}
<|\psi_{p}^{h}|\vec{p}|\psi_{m}^{e}>=\sum_{i,j=1,n}a_{j}^{p*}a_{i}^{m}<\phi_{i}^{h}|\vec{p}||\phi_{i}^{e}>.
\label{eq:eq_seven}
\end{equation}

The main contributions in eqn. (\ref{eq:eq_seven}) come from the momentum matrix elements between orbitals $\phi_{i}^{h}>$ and $\phi_{i}^{e}$ localized on the same site, i.e. when $i=j$. Using the fact that $A = <|\phi_{i}^{h}|p||\phi_{i}^{e}>$ is independent of the site $i$ by translational symmetry, we obtain:

\begin{equation}
<|\psi_{p}^{h}|\vec{p}|\psi_{m}^{e}>=\sum_{i,j=1,n}a_{j}^{p*}a_{i}^{m}<\phi_{i}^{h}|\vec{p}||\phi_{i}^{e}>.
\label{eq:eq_eight}
\end{equation}

where $\delta_{pm} = 1$ if $p = m$, $0$ otherwise. The last equality in eqn. (\ref{eq:eq_eight})) which results from the orthonormality of the eigenstates explains the selection rule $p = m$. The same conclusion could be obtained using the effective mass theory which has been extensively used in the field of semiconductor heterostructures \cite{ref30}. The second absorption band (Fig. \ref{fig:fig1}) is associated with transitions from the states $h_{p}$ to  states $(e*)$ composed of {\it }s and $p_{\sigma}$ orbitals localized on each thiophene group. These latter states have a constant energy which means that there is almost no coupling between states localized on neighbor groups. They give rise to a flat band in the case of the polymer (Fig. \ref{fig:fig5}). A more detailed comparison between theory and experiments is difficult because individual transitions cannot be identified in the experimental absorption spectra (Fig. \ref{fig:fig1}). The broadening of the peaks may result from the coupling of the electrons to vibration modes. This interpretation is supported by the presence of several equidistant replica of the lowest transition \cite{ref7}. The separation of 0.15-0.2 eV could be associated to stretching modes of the double bonds in the conjugated system. Other sources of broadening could be the interactions of the molecule with the solvent as well as some rotational disorder.

In order to investigate the possible effect of the rotational disorder on the gap, we have calculated the electronic structure of 2TV when a thiophene group is twisted with respect to the remaining of the molecule. The torsion angle is defined in Fig. \ref{fig:fig1}, the zero angle corresponding to a planar conformation. We have already mentioned that the optimized geometry of 2TV corresponds to an almost planar conformation. Because the torsion potential is very flat between -10$^{\circ}$ and +10$^{\circ}$ with variations in energy of only few meV a precise determination of the torsion angle is clearly beyond the accuracy of such calculation. However, in spite of this uncertainty on the position of the energy minimum, our calculations support that nTVs are a favorable case to get a planar conformation which allows a maximum delocalization of the electron and hole states. It is interesting to compare with other conjugated oligomers. In the case of bithiophene, a $\sim 30^{\circ}$ angle between the two rings is calculated from {\it ab initio} calculations \cite{ref31} and $\sim  26^{\circ}$ from semi-empirical restricted Hartree-Fock Austin Model 1 (AM1) \cite{ref8}. High-torsion angles are also predicted in the case of three base forms of polyaniline, i.e. leocoemeraldine, emeraldine, and pernigraniline \cite{ref10}. A planar conformation has been assumed in the case of AM1 calculations on oligo(phenylenevinylene)s \cite{ref9}, in agreement with some fluorescence studies on stilbene \cite{ref32} but in disagreement with electron diffraction experiments which give an angle of $\sim$30$^{\circ}$ \cite{ref33}. 

The increase in the total energy of 2TV is 0.33 eV at 90$^{\circ}$ but is only 46 meV at 30$^{\circ}$. Thus, at room temperature, the thermal fluctuations can probably induce angle variations of the order of 20$^{\circ}$. Obviously the geometry is also controlled by the interactions of the molecule with its environment: for example, planar conformation of oligomers is usually favored in the solid state because of packing effects \cite{ref34}. Fig. \ref{fig:fig6} shows that the gap energy increases with the torsion angle (tight binding calculations predict a smaller increase than LDA). This is the expected result since the coupling between $\pi$ orbitals of opposite groups is decreasing. A similar evolution of the gap has been obtained for biphenyl, bipyrrole and bithiophene \cite{ref35}. However, the influence of the angle fluctuations on the optical properties is small: a torsion angle of 30$^{\circ}$ gives a shift of only $\sim$ 0.2 eV of the gap of 2TV (Fig. \ref{fig:fig6}). 

Cyclic Voltammetric experiments have also been performed and analyzed in ref. \cite{ref7}. The oxidation and reduction potentials of various {\it n}TVs ($n=4 $to $12$) are plotted versus $1/N_{c}$ on (Fig. \ref{fig:fig7}). We compare these potentials to the ionization potentials and the electron affinities calculated for the free molecules. They are obtained in LDA by computing the total energy of the molecules in the +1, 0, -1 charge states. The calculated ionization potential of thiophene is in good agreement with the experimental one measured in the gas-phase \cite{ref36}. The experimental oxidation and reduction potentials are given with respect to the Ag/AgCl reference. In order to make quantitative comparisons to free-molecule values, we follow the procedure commonly adopted in the literature \cite{ref23,ref37}: we take the zero of the Ag/AgCl scale to be 4.7 eV with respect to vacuum. The calculated ionization potentials and electron affinities are approximately linear functions of $1/N_{c}$. This behavior mainly reflects the variation of the Coulomb energy associated with the ionization of the molecule. The measured oxidation and reduction potentials have a smaller dependence on $1/N_{c}$. We attribute the difference between experimental and theoretical potentials to solvation effects. To check this, we have performed LDA calculations including the Conductor-like Screening Model (COSMO) \cite{ref12,ref38}. COSMO is a continuum solvation model where the solute molecule forms a cavity within the dielectric continuum of permittivity $\epsilon$ that represents the solvent. It provides a good approximation of the electrostatic contribution to the solvation energy \cite{ref12,ref38}. Fig. \ref{fig:fig7} shows that theory and experiments agree well when solvation effects are included. 

In conclusion, we have presented theoretical studies on thienylenevinylene oligomers. LDA and tight binding calculations gives results in agreement with the experiments for the gap versus size, for the optical properties and for the electrochemical redox potentials. We propose a simple model of the electronic structure and of the optical spectra of the oligomers which can be particularly useful for further studies. We predict a small effect of the rotational disorder on the electronic structure and we obtain an efficient delocalization of the electronic states close to the gap. Combined theoretical and experimental results support that {\it n}TVs are good candidates to be used as molecular wires.

\begin{acknowledgments}
The "Institut d'Electronique et de Micro\'electronique du Nord" and "Ing\'enierie Mol\'eculaire et Mat\'eriaux Organiques" are respectively "Unit\'e Mixte 9929 et 6501 du Centre National de la Recherche Scientifique".
\end{acknowledgments}

\section{appendix}
The tight binding eigenfunctions $\psi_{k}$ of energy $E_{k}$ are written as linear combinations of atomic orbitals ~ i~ ($i$ is the atomic index and $\alpha$ is the orbital index $s$, $x$, $y$ or $z$):

\begin{equation}
\psi_{k}=\sum_{i, \alpha} a_{i \alpha}^{k} \chi_{i \alpha}.
\end{equation}

From this we calculate the number of electrons M$_{i}$ on the $i$th atom as:

\begin{equation}
M_{i} =\sum_{k,\alpha}|a_{i \alpha}^{k}|^{2},
\end{equation}

where the sum over $k$ is restricted to occupied states (including spin). The atomic population M$_{i}$ are in general different from the neutral-free-atom values Z$_{i}$, i.e. each atom bears a net charge. The self-consistency is incorporated in a simple manner following ref. \cite{ref18}. We assume that only the diagonal terms of the Hamiltonian matrix are charge dependent. They are written \cite{ref18} (in atomic units):

\begin{equation}
H_{i \alpha, i \alpha} =H^{0}_{i \alpha,i,\alpha}+\sum_{j}(M_{j}-Z_{j})\gamma_{ij}
\end{equation}

with

\begin{equation}
\gamma_{ij}=(R_{ij}^{2}+U^{-2})^{-1/2}
\end{equation}

R$_{ij}$ is the distance between atoms $i$ and $j$. $U$ is the intra-atomic Coulomb energy and the quantities $H^{0}_{i \alpha, i \alpha}$ define the {\it s} and {\it p} atomic levels in the molecule. The nondiagonal matrix elements are restricted to first nearest neighbor interactions and to two-centers integrals. They are written in terms of a small number of parameters following the rules of Slater and Koster \cite{ref15}. These parameters, fitted on the LDA electronic structure of small molecules, are given in \ref{tab:table1}. Following ref. \cite{ref16} they depend on the interatomic distance {\it d} as $(d_{0}/d)^{2}$ where d$_{0}$ is also given in Table 1. The intra-atomic Coulomb energy $U$ is equal to 10 eV for all the atoms.

\clearpage

\begin{table}[t]
\caption{Tight-binding parameters following the notations of Slater and Koster \cite{ref15}. All the energies are in electron-volts. The first nearest-neighbor interactions are given for an interatomic distance d$_{0}$.}
\label{tab:table1}
\begin{tabular}{cccc}
\hline\hline
Atomic&C-C&C-H&C-S\\
levels&interactions&interactions&interactions\\
\hline
Carbon&(ss$\sigma$)=-2.85&(ss$\sigma$)=-5.50&(ss$\sigma$)=-3.78\\
E$_{s}$=-14.57&          &                &                \\
E$_{p}$=-5.67 &          &                &                \\
Sulfur&(sp$\sigma$)=2.93 &(sp$\sigma$=6.90) &(sp$\sigma$)=4.80 \\
E$_{s}$=-16.77&          &                &                \\
E$_{p}$=-9.07 &          &                &                \\
Hydrogen&(pp$\sigma$)=3.70&                &(pp$\sigma$)=4.60  \\
E$_{s}$=-2.07&           &                &                \\
Sulfur     &(pp$\pi$)=-1.90 &             &(pp$\pi$)=-2.10   \\
           &d$_{0}$=1.54 \r{A}&d$_{0}$=1.07 \r{A}&d$_{0}$=1.54 \r{A}\\
\hline\hline
\end{tabular}
\vspace{5cm}
\end{table}

\clearpage

\begin{figure}
\caption{\label{fig:fig1} Molecular structure of 2TV and definition of the dihedral angle investigated in this work.}
\end{figure}

\begin{figure}
\caption{\label{fig:fig2} Experiments: UV-visible absorption spectra of various nTVs recorded in CH$_{2}$Cl$_{2}$. Theory: optical transition energies calculated in tight binding (arrows). The height of the arrows is proportional to the oscillator strengths.}
\end{figure}

\begin{figure}
\caption{\label{fig:fig3} Experiments: optical gaps of {\it n}TVs in solution ($+$) versus the number of sp$^{2}$ carbons $N_{c} = 6n-2$. The optical gap of thiophene in solution is from ref. \cite{ref22}. Theory: gaps calculated in LDA ($\circ$) and in tight binding ($\times$). The calculated gaps are shifted 1.2 eV upward compared to experimental data (scale on the right). The line is a fit of the tight binding gaps using $1.85 + \frac{14.96}{N_{c}^{1.12}}$ (in eV).}
\end{figure}

\begin{figure}
\caption{\label{fig:fig4} Electronic states calculated in LDA from thiophene to 6TV. The gap is represented by arrows.}
\end{figure}

\begin{figure}
\caption{\label{fig:fig5} Left: band structure of polythienylenevinylene calculated in tight binding (the conduction bands are not shifted). Dashed curves: fits of the highest occupied band and the lowest unoccupied band by a cosine law. The horizontal dashed lines are the electron and hole states calculated for 6TV using the simple model of a "s" band. The vertical arrows correspond to the main optical transitions. Right: some electron and hole states of 6TV.}
\end{figure}

\begin{figure}
\caption{\label{fig:fig6} Gap (continuous line: LDA, dashed line: tight binding) of 2TVS with respect to the torsion angle.}
\end{figure}

\begin{figure}
\caption{\label{fig:fig7} Electron affinities (lower +'s) and ionization potentials (upper +'s) of {\it n}TVs in the gas phase calculated in LDA versus $1/N_{c}$ where $N_{c}$ is the number of sp$^{2}$ carbons in the molecule; same quantities calculated for the molecules in the solvent ($\times$); experimental reduction potentials (lower $\diamond$'s) and oxidation potentials (upper $\diamond$'s). Experimental ionization energy of thiophene in the gas phase \cite{ref36}(black square). Solid lines: linear fit of the gas phase electron affinities and ionization potentials versus $1/N_{c}$.}
 \end{figure}

\cleardoublepage
\newpage

\includegraphics{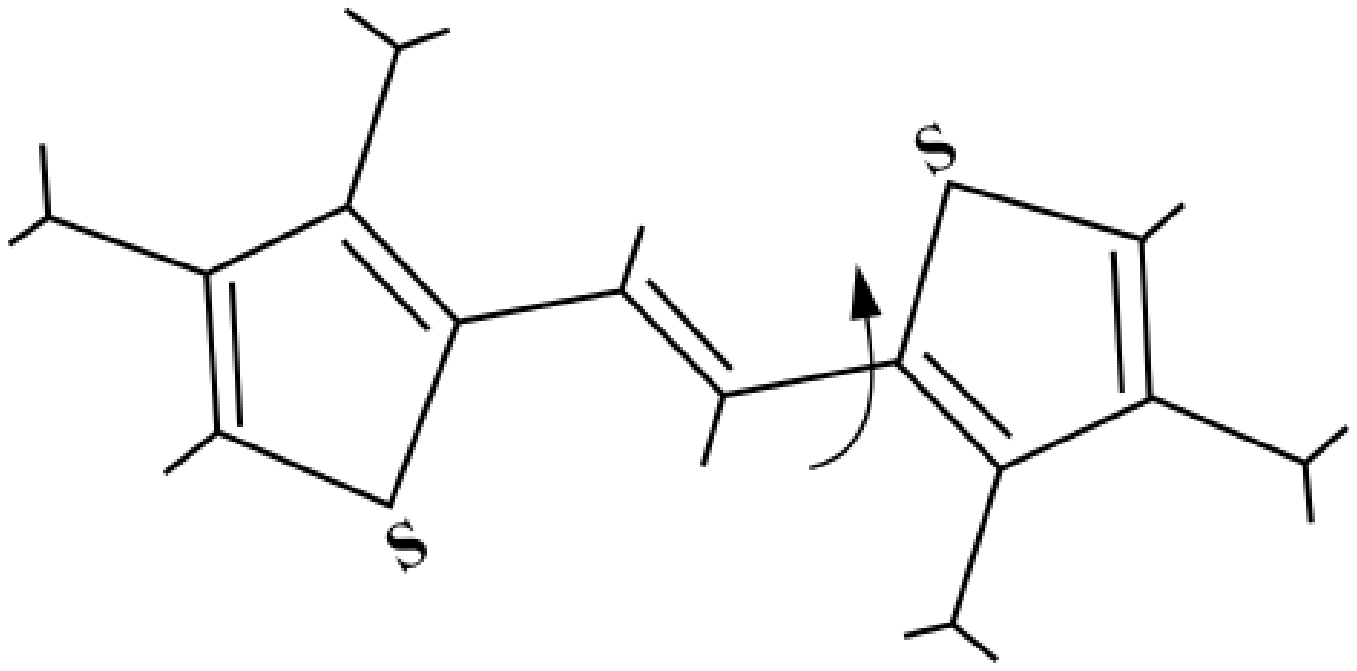}
\includegraphics{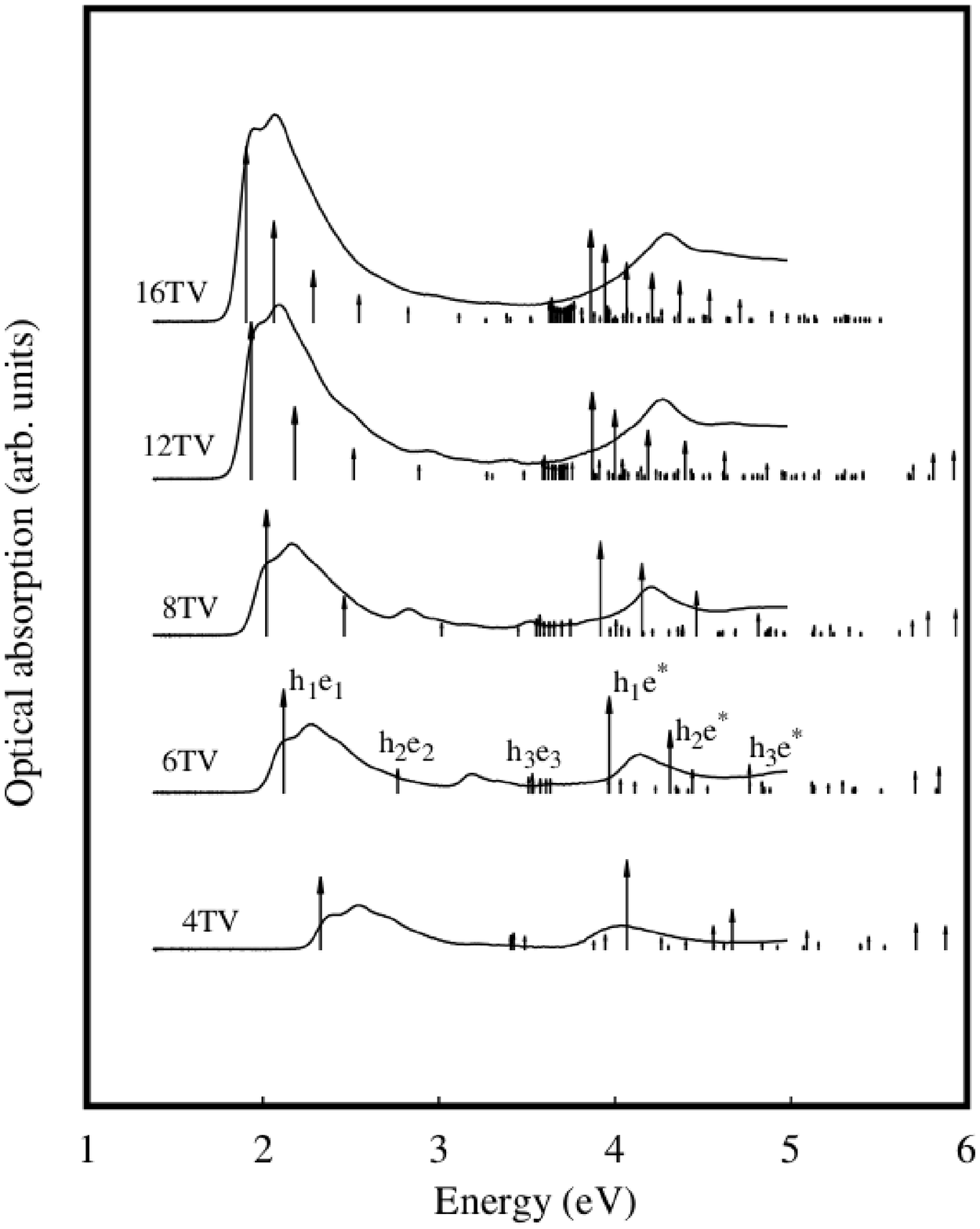}
\includegraphics{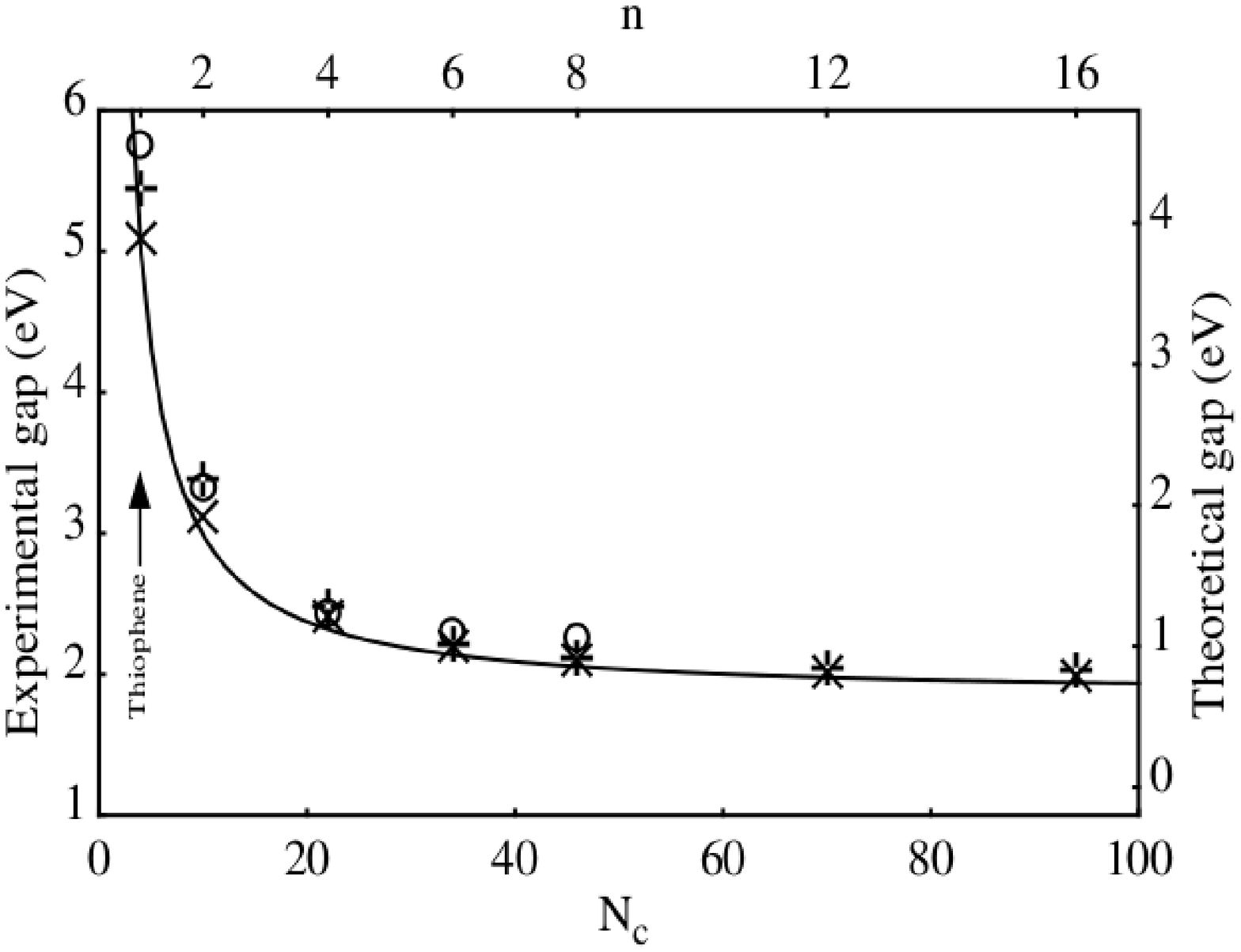}
\includegraphics{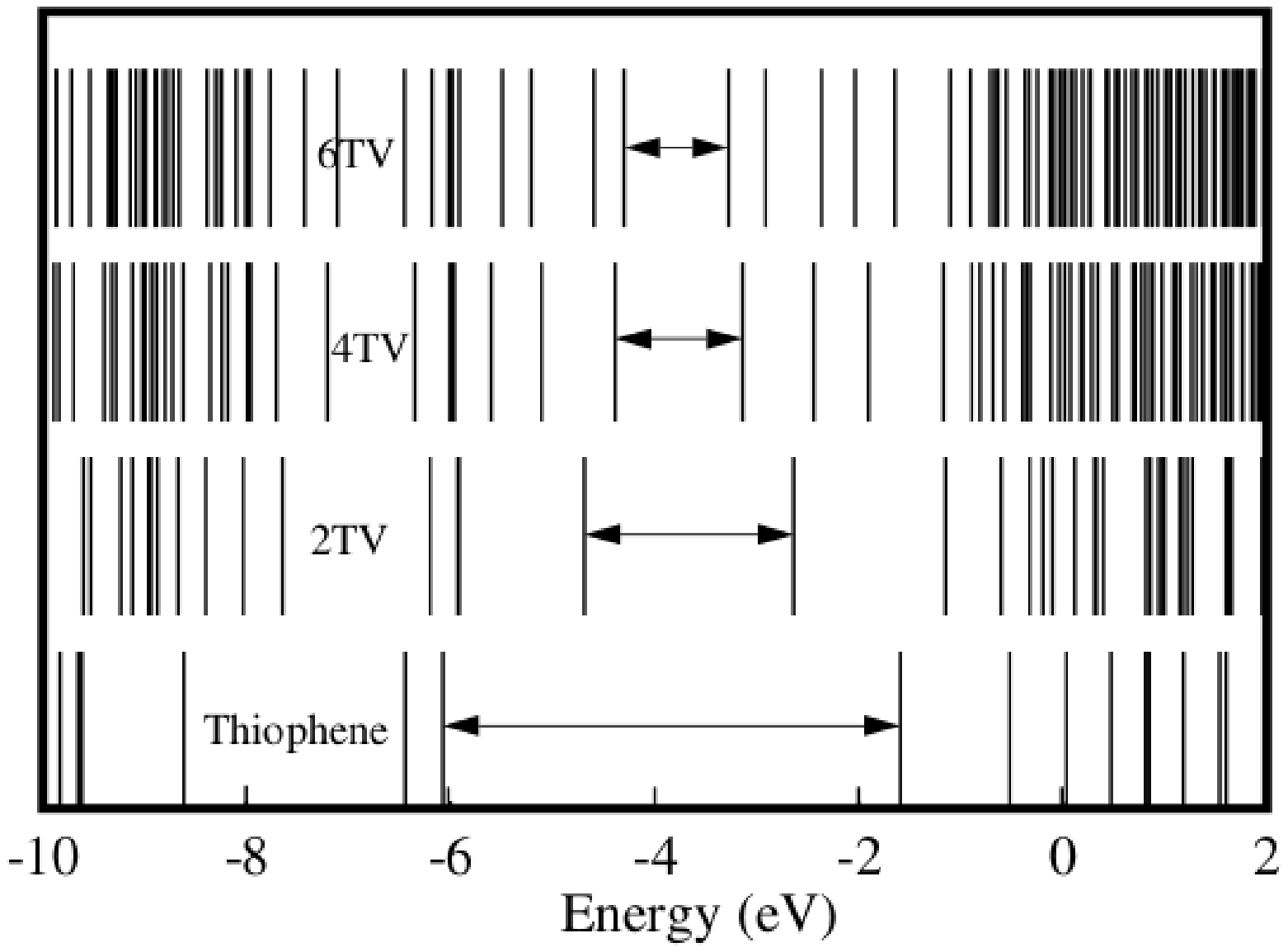}
\includegraphics{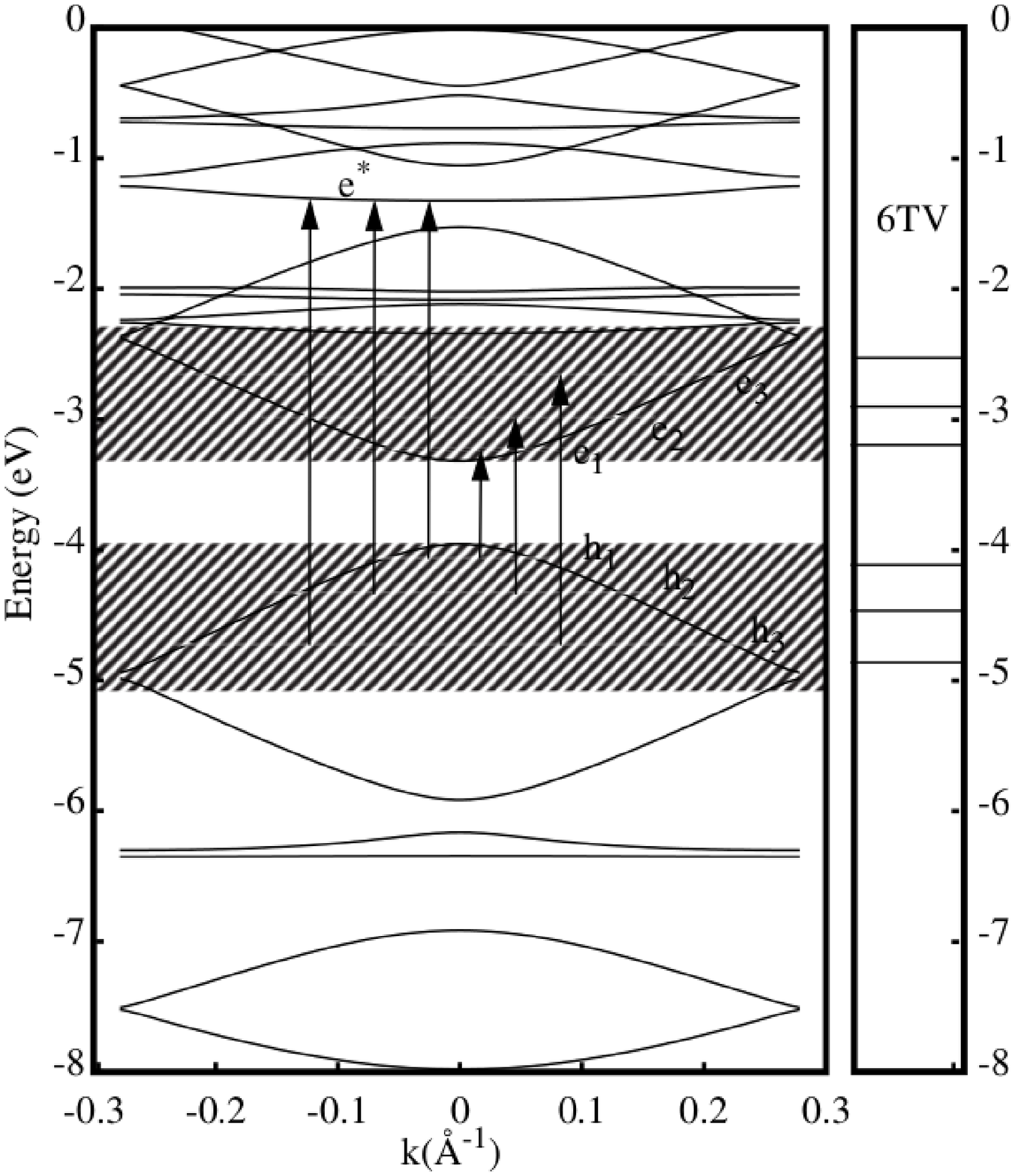}
\includegraphics{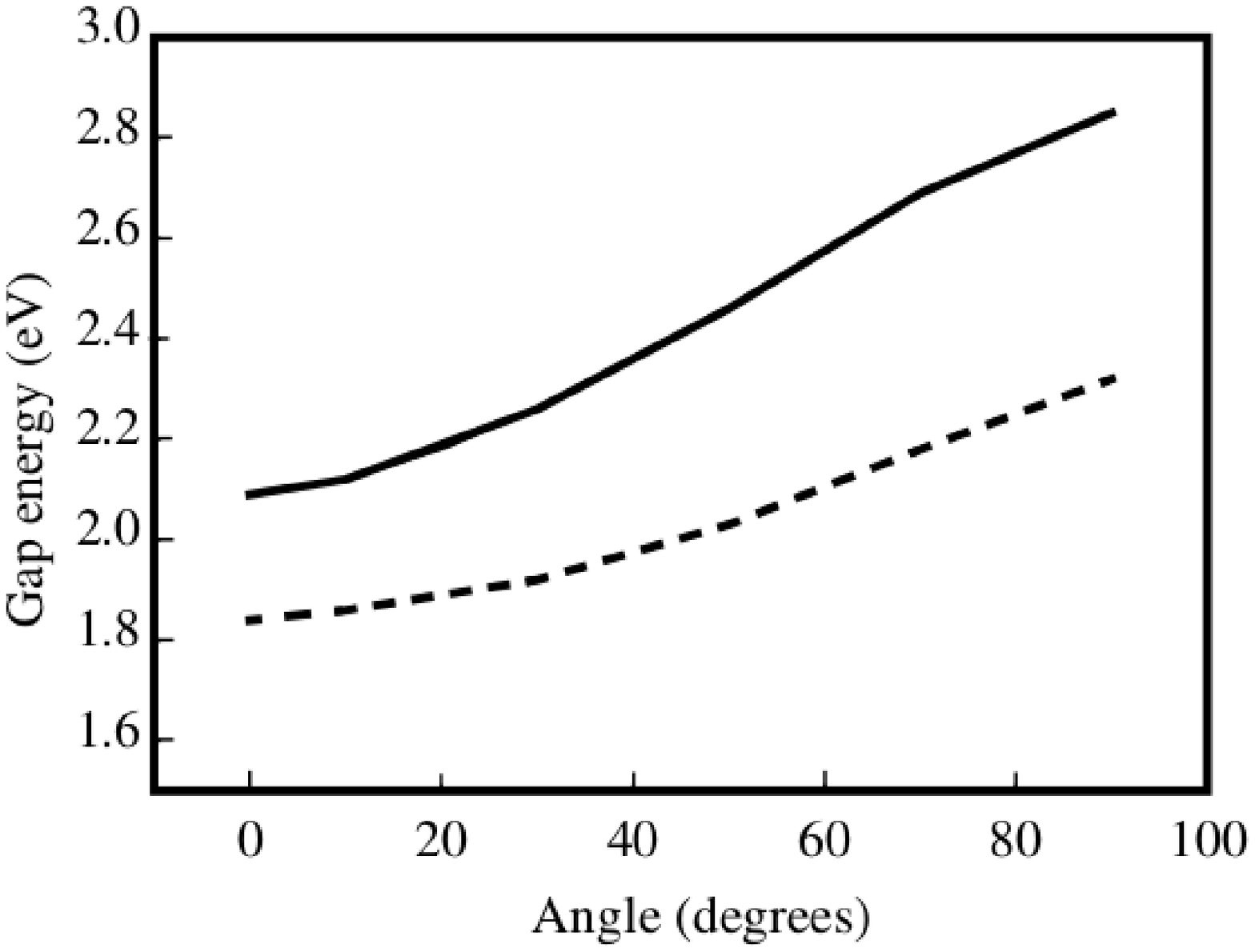}
\includegraphics{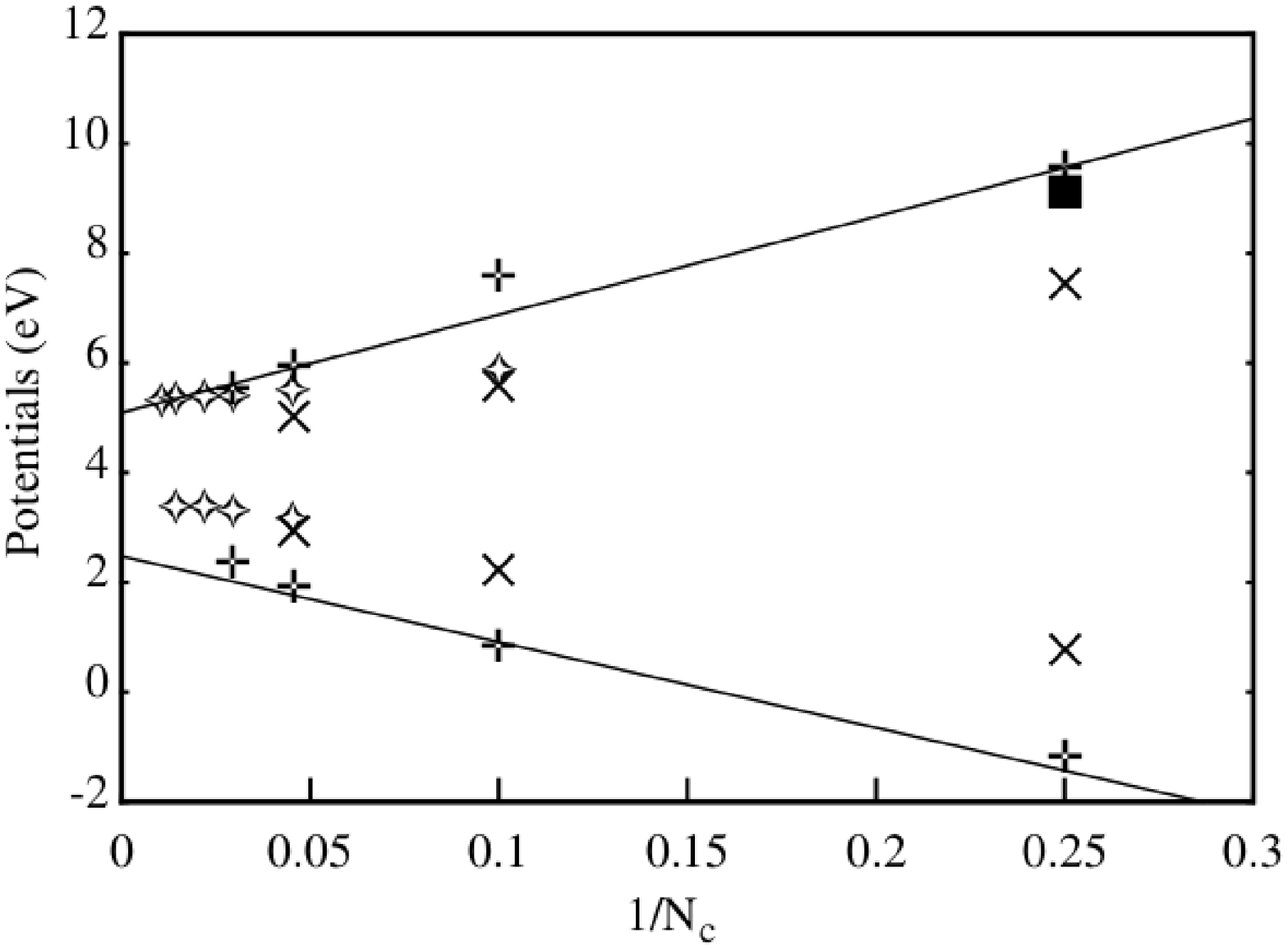}


\begin{thebibliography}{0}
\expandafter\ifx\csname natexlab\endcsname\relax\def\natexlab#1{#1}\fi
\expandafter\ifx\csname bibnamefont\endcsname\relax
  \def\bibnamefont#1{#1}\fi
\expandafter\ifx\csname bibfnamefont\endcsname\relax
  \def\bibfnamefont#1{#1}\fi
\expandafter\ifx\csname citenamefont\endcsname\relax
  \def\citenamefont#1{#1}\fi
\expandafter\ifx\csname url\endcsname\relax
  \def\url#1{\texttt{#1}}\fi
\expandafter\ifx\csname urlprefix\endcsname\relax\def\urlprefix{URL }\fi
\providecommand{\bibinfo}[2]{#2}
\providecommand{\eprint}[2][]{\url{#2}}

\end{thebibliography}


\newpage

\begin{thebibliography}{}

\bibitem{ref1} J. Gimzewski, {\it Physics World}, June 1998, 29 {\bf (1998)}.
\bibitem{ref2} D.L. Pearson, J.S. Schumm and J.M. Tour, {\it Macromolecules} 27, 2348 {\bf (1994)}.
\bibitem{ref3} R. Wu, J.S. Schumm, D.L. Pearson and J.M. Tour, {\it J.Org.Chem. } 61, 6906 {\bf (1996)}.
\bibitem{ref4} J. Roncali, {\it J.Chem.Rev}. 97, 173 {\bf (1997)}.
\bibitem{ref5} G. Kossmehl,{\it  Ber. Bunsenges. Phys. Chem.} 83, 417 {\bf (1979)}; G.Kossmehl, M. Hartel and G. Manecke,  {\it Makromol. Chem.} 131, 15 {\bf (1970)}.
\bibitem{ref6} J. Nakayama and T. Fujimori, {\it Heterocycles} 32, 991 {\bf (1991)}.
\bibitem{ref7} E. Elandaloussi, P. Fr\`ere, P. Richomme, J. Garin, and J. Roncali,{\it J. Am. Chem.S oc.} 119, 10774 {\bf (1997)}; I. Jestin, P. Fr\` ere, P. Blanchard and J. Roncali, {\it Angew. Chem. Int. Ed.} 37, 942 {\bf (1998)};I. Jestin, P. Fr\`ere, N. Mercier, E. Levillain, D. Sti\'evenard and J. Roncali, {\it J. Am. Chem. Soc.} 120,8150 {\bf (1998)}.
\bibitem{ref8} J. Cornil, D. Belijonne, and J.L. Br\'edas, {\it J. Chem. Phys.} 103, 842 {\bf (1995)}.
\bibitem{ref9} J. Cornil, D. Belijonne, and J.L. Br\'edas, {\it J. Chem. Phys.} 103, 834 {\bf (1995)}.
\bibitem{ref10} J. Libert, J. Cornil, D.A. dos Santos, and J.L. Br\'edas, {\it Phys. Rev. B} 56, 8638 {\bf (1997)}.
\bibitem{ref11} P. Hohenberg and W. Kohn, Phys.Rev. 136, 864 (1964). W. Kohn and L.J. Sham, {\it Phys. Rev.} 140, A1133 {\bf (1965)}.
\bibitem{ref12} Cerius$^{2}$ User Guide. San Diego: Molecular Simulations Inc. (1997).
\bibitem{ref13} S.J. Vosko, L. Wilk, and M. Nusair,{\it  Can. J. Phys.} 58, 1200 {\bf  (1980)}.
\bibitem{ref14} J.P. Perdew and Y. Wang, {\it Phys. Rev. B} 45, 13244 {\bf (1992)}.
\bibitem{ref15} J.C. Slater and G.F. Koster, {\it The Physical Review} 94, 1498 {\bf  (1954)}.
\bibitem{ref16} W.A. Harrison, Electronic Structure and the Properties of Solids (Freeman, San Francisco,
1980).
\bibitem{ref17} Mulliken, R.S. {\it J. Chem. Phys}. 23, 1833 {\bf (1955)}
\bibitem{ref18} M. Lannoo, {\it Phys. Rev. B} 10, 2544 {\bf (1974)}; M. Lannoo and J. Bourgoin, {\it in Point Defects in Semiconductors I}, edited by M. Cardona (Springer-Verlag, New York, 1981).
14
\bibitem{ref19} H. Burke and E.K.U. Gross, in Density Functionals: Theory and Applications, edited by D. Joubert (Springer, Berlin, 1998).
\bibitem{ref20} R.O. Jones and O. Gunnarsson, {\it Reviews of Modern Physics} 61, 689 {\bf (1989)}.
\bibitem{ref21} C. Delerue, M. Lannoo and G. Allan, unpublished.
\bibitem{ref22} A.F. Diaz, J. Crowley, J. Baryon, G.P. Gardini, and J.B. Torrance, {\it J .Electroanal. Chem.} 121, 355 {\bf (1981)}.
\bibitem{ref23} J.L. Br\'edas, R. Silbey, D.S. Boudreaux, and R.R. Chance,{\it  J. Am. Chem. Soc.} 105, 6555 {\bf (1983)}.
\bibitem{ref24} H. Meier, U. Stalmach, H. Kolshorn, {\it Acta Polymer}. 48, 379 {\bf (1997)}.
\bibitem{ref25} S. Yamada, S. Tokito, T. Tsuisui, S. Saito, {\it Chem. Commun.}, 1448 {\bf (1987)}. J. Barker, {\it Synth. Met.} 32, 43 {\bf (1989)}. H. Heckhardt, L.W. Shacklette, K.Y. Jen, R.L. Elsenbaumer, {\it J. Chem. Phys.} 91,
1303 {\bf (1989)}.
\bibitem{ref26} J.P. Proot, C. Delerue and G. Allan, {\it Appl. Phys. Lett.} 61, 1949 {\bf (1992)}.
\bibitem{ref27} A. Streitwieser, Jr., Molecular Orbital Theory (Wiley, New York 1961); C.A. Coulson, {\it Proc. R. Soc. London Ser. A} 169, 413 {\bf (1939)}; C.A. Coulson and H.C. Longuet-Higgins, {\it Proc. R. Soc. London Ser. A} 192, 16 {\bf (1947)}.
\bibitem{ref28} D. L. Dexter, in Solid State Physics, Advances in Research and Applications, edited by F. Seitz and D. Turnbull (Academic, New York), Vol. 6, p. 360 {\bf (1958)}.
\bibitem{ref29} J. Petit, G. Allan, and M. Lannoo, {\it Phys. Rev. B} 33, 8595 {\bf (1986)}.
\bibitem{ref30} G. Bastard, in Wave mechanics applied to semiconductor heterostructures (Les Editions de Physique, Les Ulis, France, 1988).
\bibitem{ref31} C. Quattrocchi, R. Lazzaroni, and J.L. Br\'edas, {\it Chem. Phys. Lett.} 208, 120 {\bf (1993)}.
\bibitem{ref32} L.H. Spangler, R. Van Zee, and T.S. Zwier, {\it J. Phys. Chem.} 91, 6077 (1987).
\bibitem{ref33} M. Troetteberg, E.B. Frantsen, F.C. Mijlhoff, and A. J. Hoekstra, {\it J. Mol. Struct.} 26, 57 (1975).
\bibitem{ref34} G. Mao, J.E. Fischer, F.E. Karasz, and M.J. Winokur, {\it J. Phys. Chem.} 98, 712 (1993).
\bibitem{ref35} J.L. Br\'edas, G.B. Street, B. Th\'emans, and J. M. Andr\'e, {\it J. Chem. Phys.} 83, 1323 (1985).
\bibitem{ref36} G. Herzberg, Electronic Spectra of Polyatomic Molecules, Van-Nostrand, New York (1976).
\bibitem{ref37} F. Lohmann, {\it Z. Naturforsch., Teil A} 22, 843 {\bf (1967)}.
\bibitem{ref38} A. Klamt and G. Schüürmann, {\it J. Chem. Soc. Perkin Trans.} 2, 799 {\bf (1993)}. J. Tomasi and M. Persico, {\it Chem. Rev.} 94, 2027 {\bf (1994)}.


\end{thebibliography}
\end{document}